 \documentclass[final,5p,times,twocolumn]{elsarticle}

\newif\ifColors    \Colorsfalse
\newif\ifShowSigns    \ShowSignstrue

\usepackage{natbib}
\biboptions{sort&compress}

\usepackage{verbatim}
\usepackage{amsmath,amsthm,mathtools,amssymb,amsfonts}
\usepackage{relsize}
\usepackage[cal=dutchcal]{mathalfa}
\usepackage{courier}

\newcommand{\mathscr}[1]{\mathcal{#1}}

\usepackage{hyperref}
\usepackage[dvipsnames,hyperref]{xcolor}
\hypersetup{
  bookmarks = true,
  hidelinks
}

\usepackage[formats]{listings}
\usepackage{enumitem}

\newcounter{bla}

\journal{Computer Physics Communications}

\newif\ifColors    \Colorsfalse
\newif\ifShowSigns    \ShowSignstrue

\hypersetup{
	pdfstartview={Fit}, pdfpagelayout={SinglePage}
}

\usepackage{mathabx}
\usepackage{accents}
\usepackage{xspace}

\DeclareMathAlphabet\urwscr{U}{urwchancal}{m}{n}%

\ifx \ii \undefined  

\DeclareMathAlphabet{\mathsfit}{\encodingdefault}{\sfdefault}{m}{sl}
\SetMathAlphabet{\mathsfit}{bold}{\encodingdefault}{\sfdefault}{bx}{sl}

\DeclareMathAlphabet{\mathesstixfrak}{U}{esstixfrak}{m}{n}

\newcommand{\rb}[1]{\left(#1\right)}

\newcommand{\cb}[1]{\left\lbrace #1 \right\rbrace}

\DeclareSymbolFont{eulerletters}{U}{eus}{m}{n}
\DeclareMathSymbol{\eulU}{\mathord}{eulerletters}{`U}
\DeclareMathSymbol{\eulC}{\mathord}{eulerletters}{`C}
\DeclareMathSymbol{\eulB}{\mathord}{eulerletters}{`B}
\DeclareMathSymbol{\eulQ}{\mathord}{eulerletters}{`Q}
\DeclareMathSymbol{\eulE}{\mathord}{eulerletters}{`E}
\DeclareMathSymbol{\eulD}{\mathord}{eulerletters}{`D}
\DeclareMathSymbol{\eulV}{\mathord}{eulerletters}{`V}
\DeclareMathSymbol{\eulW}{\mathord}{eulerletters}{`W}
\DeclareMathSymbol{\eulS}{\mathord}{eulerletters}{`S}
\DeclareMathSymbol{\eulG}{\mathord}{eulerletters}{`G}
\DeclareMathSymbol{\eulL}{\mathord}{eulerletters}{`L}
\DeclareMathSymbol{\eulK}{\mathord}{eulerletters}{`K}
\DeclareMathSymbol{\eulR}{\mathord}{eulerletters}{`R}
\DeclareMathSymbol{\eulN}{\mathord}{eulerletters}{`N}
\DeclareMathSymbol{\eulT}{\mathord}{eulerletters}{`T}
\DeclareMathSymbol{\eulO}{\mathord}{eulerletters}{`O}
\DeclareMathSymbol{\eulX}{\mathord}{eulerletters}{`X}

\DeclareSymbolFont{CMB}{OMS}{cmsy}{m}{it}
\DeclareMathSymbol{\cmA}{\mathord}{CMB}{`A}
\DeclareMathSymbol{\cmB}{\mathord}{CMB}{`B}
\DeclareMathSymbol{\cmW}{\mathord}{CMB}{`W}

\definecolor{red}{rgb}{1,0,0.1}
\definecolor{green}{rgb}{0.0,0.6,0}
\definecolor{blue}{rgb}{0.1,0.1,1}
\definecolor{orange}{rgb}{0.6,0.3,0}
\definecolor{magenta}{rgb}{0.9,0.1,1}

\newcommand{\qm}[1]{``#1"}

\newcommand\nPlusOne{$\sdim\hspace{-0.1em}+\hspace{-0.1em}1$\xspace}

\newcommand\threePlusOne{$3\hspace{-0.075em}+\hspace{-0.1em}1$\xspace}

\renewcommand\tilde[1]{\mkern1mu\widetilde{\mkern-1mu#1}}

\global\long\def\rb#1{\left(#1\right)}
 
 \global\long\def\cb#1{\left\lbrace #1 \right\rbrace }
 \global\long\def\qm#1{``#1''}

\newcommand\bSe{\begin{subequations}}
\newcommand\eSe{\end{subequations}}

\newcommand\mat[1]{
\begin{pmatrix}
#1
\end{pmatrix}
}

\newcommand\matc[1]{
  \bgroup\renewcommand{\arraystretch}{0.9}\begin{pmatrix}
	#1
  \end{pmatrix}\egroup
}

\usepackage{xparse}
\ExplSyntaxOn
\NewDocumentCommand{\mref}{m}{\quinn_mref:n {#1}}
\seq_new:N \l_quinn_mref_seq
\cs_new:Npn \quinn_mref:n #1
 {
  \seq_set_split:Nnn \l_quinn_mref_seq { , } { #1 }
  \seq_pop_right:NN \l_quinn_mref_seq \l_tmpa_tl
  ( 
  \seq_map_inline:Nn \l_quinn_mref_seq
    { \ref{##1},\nobreakspace } 
  \exp_args:NV \ref \l_tmpa_tl 
  ) 
 }
\ExplSyntaxOff

\newcommand\ii{\mathrm{i}}
\newcommand\ee{\mathrm{e}}

\newcommand\tr{\mathtt{{\scriptscriptstyle \top}}}

\newcommand\sdim{N}

\ifColors
  \newcommand\gSector[1]{{\color{blue}#1}}
  \newcommand\fSector[1]{{\color{red}#1}}
  \newcommand\hSector[1]{{\color{green}#1}}
  \newcommand\lSector[1]{{\color{BurntOrange}#1}}

\else
  \newcommand\gSector[1]{{\color{black}#1}}
  \newcommand\fSector[1]{{\color{black}#1}}
  \newcommand\hSector[1]{{\color{black}#1}}
  \newcommand\lSector[1]{{\color{black}#1}}

\fi

\newcommand\gMet{\gSector g}
\newcommand\gSp{\gSector{\gamma}}
\newcommand\gLapse{\gSector{\alpha}}
\newcommand\gShift{\gSector \beta}

\newcommand\gK{\gSector K}
\newcommand\gE{\gSector e}
\newcommand\gFE{\gSector E}

\newcommand\fMet{\fSector f}
\newcommand\fSp{\fSector{\varphi}}
\newcommand\fLapse{\fSector{\protect\tilde{\alpha}}}
\newcommand\fShift{\fSector{\protect\tilde{\beta}}}

\newcommand\fK{\fSector{\protect\tilde{K}}}
\newcommand\fE{\fSector m}

\newcommand\fEtr{\fSector{m_\mathtt{o}}}

\newcommand\fFEtr{\fSector{M_\mathtt{o}}}

\newcommand\gEA{\gSector A}

\newcommand\fEA{\fSector{\protect\tilde{A}}}

\newcommand\code[1]{$\mathtt{#1}$}

\newcommand\gBSSN[1]{{\widebar{#1}}}
\newcommand\fBSSN[1]{{\widehat{#1}}}
\newcommand\hBSSN[1]{{\protect\accentset{\circ}{{#1}}}}

\newcommand\hSpBS{\hSector{\hBSSN \chi}}

\newcommand\gconf{\gSector{\phi}}
\newcommand\fconf{\fSector{\psi}}

\newcommand\gSpBS{\gSector{\gBSSN \gamma}}

\newcommand\gABS{\gSector{\gBSSN A}}
\newcommand\gKBS{\gSector{\gBSSN K}}
\newcommand\gEBS{\gSector{\gBSSN e}}

\newcommand\gGBS{\gSector{\gBSSN \Gamma}}
\newcommand\gL{\gSector{\gBSSN \Lambda}}
\newcommand\gDG{\gSector{\bigtriangleup\gBSSN{\Gamma}}}
\newcommand\gDback{\gSector{\gBSSN{D}_\mathtt{B}}}
\newcommand\gGbackBS{\gSector{\gBSSN{\Gamma}_\mathtt{B}}}
\newcommand\gRback{\gSector{\gBSSN{R}_\mathtt{B}}}
\newcommand\gRBS{\gSector{\gBSSN{\eulR}}}

\newcommand\fSpBS{\fSector{\fBSSN \varphi}}

\newcommand\fABS{\fSector{\fBSSN A}}
\newcommand\fKBS{\fSector{\fBSSN K}}
\newcommand\fEBS{\fSector{\fBSSN m}}

\newcommand\fEtrBS{\fSector{\fBSSN{m}_\mathrm{o}}}

\newcommand\fGBS{\fSector{\fBSSN \Gamma}}
\newcommand\fL{\fSector{\fBSSN \Lambda}}
\newcommand\fDG{\fSector{\bigtriangleup\fBSSN{\Gamma}}}
\newcommand\fDback{\fSector{\fBSSN{D}_\mathtt{B}}}
\newcommand\fGbackBS{\fSector{\fBSSN{\Gamma}_\mathtt{B}}}
\newcommand\fRback{\fSector{\fBSSN{R}_\mathtt{B}}}
\newcommand\fRBS{\fSector{\fBSSN{\eulR}}}

\newcommand\hMet{\hSector h}
\newcommand\hSp{\hSector{\chi}}

\newcommand\hShift{\hSector q}

\newcommand\sLs{\lSector{\boldsymbol{\Lambda}_\mathrm{s}}}
\newcommand\sLt{\lSector{\lambda}}

\newcommand\sLp{\lSector{\boldsymbol{\mathrm{p}}}}
\newcommand\sLw{\lSector{\boldsymbol{\mathrm{w}}}}
\newcommand\sRs{\lSector{\boldsymbol{\mathrm{R}}}}
\newcommand\sRbar{\lSector{\boldsymbol{\mathrm{R}}_\mathrm{o}}}

\newcommand\sEta{\lSector{\boldsymbol{\mathrm{\delta}}}}
\newcommand\Eta{\lSector{\boldsymbol{\mathrm{\eta}}}}

\ifColors

\else

\fi

\DeclareMathVersion{normal2}

\begin{document}

\begin{frontmatter}



\title{$\mathtt{bimEX}$: A Mathematica package for exact computations in \threePlusOne bimetric relativity}


\author[a]{Francesco Torsello\corref{author}}

\cortext[author] {Corresponding author.\\\textit{E-mail address:} francesco.torsello@fysik.su.se}
\address[a]{Department of Physics \& The Oskar Klein Centre, \\
Stockholm University, AlbaNova University Center, SE-106 91 Stockholm, Sweden}

\begin{abstract}




We present \code{bimEX}, a Mathematica package for exact computations in 3$+$1 bimetric relativity. It is based on the \code{xAct} bundle, which can handle computations involving both abstract tensors and their components. In this communication, we refer to the latter case as concrete computations. The package consists of two main parts. The first part involves the abstract tensors, and focuses on how to deal with multiple metrics in \code{xAct}. The second part takes an ansatz for the primary variables in a chart as the input, and returns the covariant BSSN bimetric equations in components in that chart. Several functions are implemented to make this process as fast and user-friendly as possible. The package has been used and tested extensively in spherical symmetry and was the workhorse in obtaining the bimetric covariant BSSN equations and reproducing the bimetric \threePlusOne equations in the spherical polar chart.

\end{abstract}


\end{frontmatter}



\noindent{\bf PROGRAM SUMMARY.} \\

\begin{small}
\noindent
{\em Program Title:} \code{bimEX}                                       \\
{\em Program Files doi:} \url{http://dx.doi.org/10.17632/2s5d7csc9w.1} \\
{\em Licensing provisions:} GNU General Public License 3.0 (GPL)                                   \\
{\em Programming language:} \code{Mathematica}                                   \\
{\em Supplementary material:}  \\
\begin{enumerate}[topsep=0mm]
	\item README file, containing instructions about how to use the working example.
	\item Working example, constituted by the notebooks:
	\begin{enumerate}
		\item bimEX\textunderscore Working\textunderscore Example.nb
		\item bimEX\textunderscore Decomposition\textunderscore Lists\textunderscore Loader.nb
		\item bimEX\textunderscore Decomposition\textunderscore xAct\textunderscore Loader.nb
	\end{enumerate}
\end{enumerate}
{\em Nature of problem (approx. 50-250 words):}\\
  Writing the bimetric covariant BSSN equations in any desired ansatz and chart. \\
{\em Solution method (approx. 50-250 words):}\\
  Definition of functions within the \code{Mathematica} package \code{xAct}, which computes all the components of the defined abstract tensors and reduce the abstract tensors to their representation in components. \\
{\em Additional comments including Restrictions and Unusual features (approx. 50-250 words):}\\
GitHub repository at \href{https://github.com/nubirel/bimEX}{https://github.com/nubirel/bimEX} \\
   \\

\end{small}

\section{Introduction}
\label{sec:introduction}

\subsection{Motivation and general description}

The Hassan--Rosen bimetric theory, or bimetric relativity (BR), is a nonlinear theory of interacting massless and massive spin-2 fields \cite{Hassan:2011zd,Hassan:2011ea,HassanKocic2018,HassanLundkvist2018}. As a theory of modified gravity, it has a rich phenomenology, and its spectrum of solutions contain both the general relativity (GR) solutions and novel solutions \cite{doi:10.1142/S0218271814430020,doi:10.1063/1.5100027}. We refer the reader to \cite{Schmidt_May_2016} for a review on the theory, to \cite{Kocic:2017hve} for a more recent non-GR exact solution of the theory, and to \cite{10.1088/1361-6382/ab4f9b} for a recent work affirming the compatibility of the theory with local tests of gravity. However, the number of non-GR solutions is modest, at present. This is due to the difficulties in solving the bimetric field equations (BFE), both analytically and numerically. Most of the non-GR solutions, see, e.g., \cite{Comelli:2011wq,PhysRevD.85.124043,Brito:2013xaa,PhysRevD.96.064003} and \cite{vonStrauss:2011mq}, have been obtained by integrating the BFE numerically after reducing them to ordinary differential equations.

In the search for solutions describing more realistic physical systems, e.g., spherically symmetric vacuum and non-vacuum solutions with non-trivial dynamics, one has to deal with a system of partial differential equations (PDEs). In GR, the Einstein field equations are also PDEs in the general case, and their numerical integration needs a recasting as a well-posed Cauchy problem. See \cite{Choquet-Bruhat:2014hta} for a review on the history of the Cauchy problem in GR. The same thing holds in BR.

The recasting of the BFE as a Cauchy problem was established in \cite{Kocic:2018ddp}. However, this recasting does not result in a well-posed formulation. For this reason, following the road suggested by numerical relativity, it is desirable to recast the BFE in the covariant BSSN formalism (see \cite{PhysRevD.52.5428,PhysRevD.59.024007} and \cite{baumgarte2010numerical} for the BSSN formulation of the Einstein field equations), which is well-posed in GR if one chooses the standard gauge and satisfies other technical conditions \cite{PhysRevD.66.064002,PhysRevD.70.104004,PhysRevD.79.104029}. In \cite{Torsello_2019}, the covariant BSSN formulation of the BFE is computed. Unfortunately, due to the particular interaction between the metrics, the well-posedness of the bimetric covariant BSSN has not been established yet.

Having found the bimetric covariant BSSN formulation, one would like to be able to compute it in any desired chart. This is the first step towards the numerical integration of the equations and the attainment of solutions to the BFE describing sensible bimetric physical systems, e.g., nonlinear bimetric gravitational collapse. \code{bimEX} was developed with this goal in mind, and was the workhorse in obtaining the results in \cite{Torsello_2019,Torsello_2019b}, and reproducing the results in \cite{Kocic:2018ddp}. The name of the package is an acronym for ``bimetric exact computations". The last part of the name, \code{EX}, can be interpreted as ``ex" in \textbf{ex}act, or ``e" in \textbf{e}xact and ``cs"($=$x) from \textbf{c}omputation\textbf{s}.

In this communication we describe \code{bimEX}, developing a working framework for \code{Mathematica} 11.0 \cite{Mathematica} or later, and \code{xAct} 1.1.3 \cite{xAct} or later, to perform computations in \threePlusOne bimetric relativity. \code{xAct} already provides an excellent framework to handle both abstract and concrete computations; with \qm{concrete} equations, we mean equations written in components in some chart. \code{bimEX} adds to it the definitions of several bimetric geometrical objects defined in \cite{Kocic:2018ddp} and \cite{Torsello_2019}, the definitions of the bimetric covariant BSSN constraint and evolutions equations introduced in \cite{Torsello_2019} and the definitions of several ready-to-use functions that allow the user to obtain the concrete equations starting from the abstract ones, in any given chart. For example, by using \code{bimEX}, the user can call only one function to write an abstract equation into components, without the need to explicitly call the \code{xAct} function \code{ToBasis}, \code{TraceBasisDummy} and the others. The latter are used in the background by \code{bimEX}. Also, another \code{bimEX} function takes the chosen ansatz on the primary variables---introduced in the next section---as its input and gives the bimetric decomposition as output, without the need for the user to explicitly write all the needed formulas defined in \cite{Kocic:2018ddp,Torsello_2019}.

All the commands and options in \code{bimEX} are documented through usage messages in \code{Mathematica}, i.e., by typing \code{?\langle name\; of\; the\; function/option\rangle} after loading the package.

\subsection{Bimetric relativity and its primary variables}

In this section, we very briefly introduce the features of bimetric relativity needed to understand the structure of \code{bimEX}, following the notation in \cite{Torsello_2019}.

In bimetric relativity, the two spin-2 fields are described by two metrics, $\gMet$ and $\fMet$. Note, however, that the metrics do not coincide with the mass eigenstates of the theory \cite{Hassan2013}. It is possible to write each one of the two metrics in terms of dynamical and kinematical variables as \cite{doi:10.1098/rspa.1958.0142,PhysRev.114.924,Arnowitt2008} (see also \cite{baumgarte2010numerical,gourgoulhon20123+1}),
\begin{subequations}
\label{eq:metrics}
\begin{align}
	\gMet	&=	\mat{
						-\gLapse^2+\gSp_{k\ell}\gShift^k\gShift^\ell 	& \gSp_{i\ell}\gShift^\ell \\
						\gSp_{j\ell}\gShift^\ell									&  \gSp_{ij}
					}, \\
	\fMet		&=	\mat{
						-\fLapse^2+\fSp_{k\ell}\fShift^k\fShift^\ell 	& \fSp_{i\ell}\fShift^\ell \\
						\fSp_{j\ell}\fShift^\ell									&  \fSp_{ij}
					}.
\end{align}
\end{subequations}
Here, the lapse functions $\gLapse,\fLapse$ and the shift vectors $\gShift,\fShift$ \cite{Wheeler:1964qna} are kinematical variables, as can be shown by performing the Hamiltonian analysis of the theory \cite{Hassan:2011zd,HassanLundkvist2018}. The spatial metrics $\gSp_{ij},\fSp_{ij}$ are instead dynamical, and are the induced metrics on a common spacelike hypersurface $\Sigma _t$ \cite{baumgarte2010numerical,gourgoulhon20123+1,Hassan:2011zd}. When writing the BFE in terms of these variables, one obtains a set of constraint equations---in the usual Hamiltonian sense---and a set of evolution equations. These equations can be written in different forms, one of them being the so-called standard \nPlusOne decomposition \cite{baumgarte2010numerical,gourgoulhon20123+1}. The bimetric standard \nPlusOne decomposition was computed in \cite{Kocic:2018ddp}.

\mathversion{normal2}
In the standard \nPlusOne decomposition, both in GR and in BR, the evolution equations are written in such a way that they contain only first-order time derivatives. In order to do that, one has to promote the time derivative of the metric to be a dynamical variable. This is done by introducing the extrinsic curvatures \cite{baumgarte2010numerical,gourgoulhon20123+1},
\begin{align}
	\gK_{ij} 	\coloneqq -\dfrac{1}{2}\mathcal{L}_{\gSector{n}}\,\gSp_{ij}, \quad
	\fK_{ij} 		\coloneqq -\dfrac{1}{2}\mathcal{L}_{\fSector{\tilde{n}}}\,\fSp_{ij},
\end{align}
where $\mathcal{L}_X$ is the Lie derivative along the vector field $X$, $\gSector{n}=(\partial_t-\gShift)/\gLapse$ and $\fSector{\tilde{n}}=(\partial_t-\fShift)/\fLapse$ are the two normal vectors to the spacelike hypersurface $\Sigma _t$, with respect to $\gMet$ and $\fMet$.

In the bimetric parametrization established in \cite{HassanKocic2018,Kocic:2018ddp}, it is made clear that, given any two metrics in \eqref{eq:metrics}, the real square root $(\gMet^{-1}\fMet)^{1/2}$ does not necessarily exist, although necessary to be able to write down the theory, since the interaction potential between the metrics depends on it.
\mathversion{normal}

A necessary and sufficient condition for the real square root to exist is that it is possible to find a Lorentz transformation $L=\Lambda\, R$, with $\Lambda $ boost and $R$ spatial rotation, such that the geometric mean of the two metrics,
\begin{align}
\label{eq:symcond}
	\hMet\coloneqq \gMet\operatorname{\#}\fMet=\gFE^\tr\Eta L\fFEtr=\hMet^\tr,
\end{align}
exists \cite{Deffayet2013,HassanKocic2018}. Here $\gFE,\fFEtr$ are the vielbeins of $\gMet, \fMet$, respectively, and $\Eta$ is the Minkowski metric.

The Lorentz transformation $L$ can be written \cite[Sec.~1.6]{meinrenken2013clifford},
\begin{align}
	L=\mat{\sLt & \sLp^\tr\sEta \\
				\sLp & \sLs}\mat{1 & 0 \\
										 0 & \sRs}=\mat{\sLt & \sLp^\tr\sEta\sRs \\
										 						\sLp & \sLs\sRs},
\end{align}
with $\sLt=\sqrt{1+\sLp^\tr\sEta\sLp}$ being the Lorentz factor of the boost, and $\sLp$ is a real spatial vector in the Lorentz frame, called \qm{separation parameter}, with components $\sLp^\textbf{a}=\sinh (\sLw^\textbf{a})$, with $\sLw^\textbf{a}$ rapidities of the Lorentz boost. The condition \eqref{eq:symcond}, in the $\sdim+1$ formalism, translates into a condition on the shifts of the two metrics, and another condition on the spatial part $\sLs\sRs$ of $L$ \cite{MikicaB.794295}. The first condition tells us that the two shifts are related,
\begin{align}
\label{eq:shiftsymcond}
	\gShift = \hShift +\dfrac{\gLapse}{\sLt}\gE^{-1}\sLp, \qquad \fShift = \hShift -\dfrac{\fLapse}{\sLt}\fE^{-1}\sLp,
\end{align}
where $\hShift$ is the shift vector of $\hMet$. The second condition requires the spatial part $\hSp$ oh $\hMet$ to be symmetric,
\begin{align}
\label{eq:spsymcond}
	\hSp=\gE^\tr\sEta\sLs\sRs\fEtr=(\gE^\tr\sEta\sLs\sRs\fEtr)^\tr=\hSp^\tr,
\end{align}
where $\gE,\fEtr,\sEta$ are the spatial parts of $\gFE,\fFEtr,\Eta$, respectively.
The Lorentz transformation $L$ such that the two conditions \eqref{eq:shiftsymcond} and \eqref{eq:spsymcond} hold, is found in two steps. First, one expresses $\sRs$ in terms of $\sLs,\gE,\fEtr$; second, one determines $\sLs$ by solving one of the constraints.\footnote{If an evolution equation for $\sLs$ is known, as in the case of spherical symmetry, one can also specify the value of $\sLs$ on the initial hypersurface and evolve it in time, without the need to solve it from one of the constraints.} Here we focus on the first step, i.e., the determination of $\sRs$. The expression of $\sRs$ is found to be \cite{MikicaB.794295}
\begin{subequations}
\label{eq:orth}
\begin{align}
	\sRs		&=(\sEta^{-1}\sRbar^\tr\sEta\sRbar)^{1/2}\sRbar^{-1},\label{eq:R} \\
	\sRbar	&\coloneqq\sEta^{-1}(\fEtr\gE^{-1})^\tr\sEta\sLs,
\end{align}
\end{subequations}
which is the polar decomposition of $\sRbar^{-1}$. Since the polar decomposition always exists, it is always possible to write $\sRs$ in these terms. In addition, $\sRbar^{-1}$ being clearly invertible, it follows that $\sEta^{-1}\sRbar^\tr\sEta\sRbar$ is strictly positive definite \cite[Sec.~2.5]{hall2015lie}. This means that the computation of $\sRs$ reduces to the computation of the square root of a $3\times 3$ symmetric positive definite matrix. This is the first step in computing the bimetric \threePlusOne and BSSN decomposition, and all the successive steps depend on it.

The recasting of the standard \nPlusOne equations in the covariant BSSN formalism, introduces new dynamical variables \cite{PhysRevD.52.5428,PhysRevD.59.024007,baumgarte2010numerical,PhysRevD.79.104029,Torsello_2019}. Following the notation in \cite{Torsello_2019}, they are the conformal metrics,
\begin{subequations}
\label{eq:confmetricscov}
	\begin{alignat}{3}
		\gSpBS_{ij}	&\coloneqq \ee^{-4\gconf}\gSp_{ij},\qquad  \gSpBS^{ij} \  &\coloneqq \ee^{4\gconf}\gSp^{ij},	\\
		\fSpBS_{ij}		&\coloneqq \ee^{-4\fconf}\fSp_{ij},\qquad  \fSpBS^{ij}		&\coloneqq \ee^{4\fconf}\fSp^{ij},
	\end{alignat}
\end{subequations}
the conformal extrinsic curvatures and traces,
\begin{subequations}
\label{eq:confAcov}
\begin{align}
	\gABS_{ij}	&\coloneqq\ee^{-4\gconf}\gEA_{ij}=\ee^{-4\gconf}\rb{\gK_{ij}-\dfrac{1}{3}\gSp_{ij}\gK+\dfrac{1}{3}\gSp_{ij}\gABS}, \\
	\fABS_{ij}	&\coloneqq\ee^{-4\fconf}\fEA_{ij}=\ee^{-4\fconf}\rb{\fK_{ij}-\dfrac{1}{3}\fSp_{ij}\fK+\dfrac{1}{3}\fSp_{ij}\fABS}, \\
	\gKBS		&= \gK-\gABS , \qquad    \fKBS = \fK-\fABS.
\end{align}
\end{subequations}
and the conformal connections,
\begin{subequations}
\begin{align}
\label{eq:confconcov}
	\gL^i \coloneqq \gSp^{jk}\gDG^i_{jk}=\gSp^{jk}\rb{\gGBS^i_{jk}-\gGbackBS^i_{jk}}, \\
	 \fL^i \coloneqq \fSp^{jk}\fDG^i_{jk}=\fSp^{jk}\rb{\fGBS^i_{jk}-\fGbackBS^i_{jk}},
\end{align}
\end{subequations}
where the background connections $\gGbackBS^{i}_{jk},\fGbackBS^{i}_{jk}$ are arbitrary but time-independent\footnote{The assumption of time-independency is due to the fact that the arbitrary connections do not have to fulfill any evolution equation. In bimetric relativity, there is the possibility to set $\hSp$ as the background geometry for $\gSp$ and $\fSp$, hence the assumption of time-independency can be relaxed. See \cite{Torsello_2019b} for more details.}, possibly arising as the compatible connections for the background metrics. We also define the conformal mean metric $\hSpBS$,
\begin{alignat}{3}
\label{eq:confmean metric}
		\hSpBS_{ij}	&\coloneqq \ee^{-2\rb{\gconf+\fconf}}\hSp_{ij},
	\end{alignat}
which is not a dynamical variable. For the bimetric covariant BSSN equations, we refer the reader to \cite{Torsello_2019}.

We are finally in the position to state what are the primary variables needed to be specified in order to compute the bimetric covariant BSSN decomposition,
\begin{align}
\label{eq:primaryvariables}
	\gconf, \fconf, \gEBS^\textbf{a}{}_{i},\fEtrBS^\textbf{a}{}_{i},\gABS^i{}_{j},\fABS^i{}_{j},\gL^i,\fL^i, \sLp^\textbf{a}, \hShift^i, \gGbackBS^{i}_{jk},\fGbackBS^{i}_{jk},
\end{align}
where $\gEBS,\fEBS$ are the conformal vielbeins,
\begin{align}
\label{eq:conftetrads}
	\gEBS=\ee ^{-2\gconf}\gE, \quad \fEBS=\ee ^{-2\fconf}\fE.
\end{align}
These are the variables used to make an ansatz, in the covariant BSSN formalism. Once they are known, the entire decomposition defined in \cite{Kocic:2018ddp,Torsello_2019} can be computed.\footnote{Note that choosing an ansatz for $\sLp^\textbf{a}$ and $\hShift^i$ is equivalent to choosing an ansatz for $\gShift^i$ and $\fShift^i$.}

\section{The abstract equations}
\label{sec:abstract}

\subsection{Basic definitions in \code{xAct}}

First, \code{bimEX} loads the \code{xAct} bundle, after which the three dimensional spacelike hypersurface $\Sigma_t$ with abstract indices \code{\{ i,j,k,q,r,s \}} is defined,
\begin{lstlisting}[caption={The definition of the spacelike hypersurface.},label={lst:hypersurface}]
	 DefManifold[|*$\mathtt{\Sigma}$*|t, 3, {i, j, k, q, r, s}]
\end{lstlisting}
The time coordinate is not defined on this manifold, but we need our objects to depend on it. For this reason, \code{bimEX} defines a parameter \code{t},
\begin{lstlisting}[caption={The definition of the time parameter.},label={lst:t}]
	 DefParameter[t]
\end{lstlisting}
All the abstract tensors defined in \code{bimEX} depend on this parameter. The time derivatives are performed using the \code{xAct} built-in function \code{ParamD[\langle parameter\rangle][\langle expression\rangle]}. Since the evolution equations in the covariant BSSN formalism contain the differential operators $\partial_t-\eulL_\gShift$ and $\partial_t-\eulL_\fShift$, we also define the parameters \code{g\digamma} and \code{f\digamma}. A third parameter \code{h\digamma} is also defined to represent the operator $\partial_t-\eulL_\hShift$. All the defined abstract tensors depend on these three parameters as well. In this way, \code{bimEX} provides a way to represent the differential operators $\partial_t-\eulL_\gShift,\partial_t-\eulL_\fShift,\partial_t-\eulL_\hShift$ in a simple way. Their explicit form must be provided by the user, when needed. These \code{xAct} representations should be considered as placeholders, but their utility in the abstract computations relies on the fact that the properties of a derivative operator, e.g., the Leibniz rule, are automatically implemented for them by \code{xAct}.

Next, \code{bimEX} defines all the abstract tensors representing the geometrical objects defined in \cite{Kocic:2018ddp,Torsello_2019}. We note that \code{bimEX} defines six metrics; the three spatial metrics $\gSp,\fSp,\hSp$ and their conformally related metrics $\gSpBS,\fSpBS,\hSpBS$. An important point to stress is that the abstract tensors representing the vielbeins of the metrics in \code{xAct}, have only spatial indices, rather than a spatial and a Lorentz index. This can potentially lead to confusion, since one cannot distinguish between the vielbein $\gE^\textbf{a}{}_i$ and its inverse $\gE^i{}_\textbf{a}$, with $\textbf{a}$ Lorentz index and $i$ spatial index. For this reason, the vielbein and the inverse vielbein are two different abstract tensors in \code{bimEX},
\begin{lstlisting}[label={lst:vielbeins},caption={The unambiguous definitions of the vielbein and its inverse.}]
	DefTensor[e[i, -j], {|*$\Sigma$*|t, t, g|*$\digamma$*|, f|*$\digamma$*|, h|*$\digamma$*|}]
	DefTensor[Inve[i, -j], {|*$\Sigma$*|t, t, g|*$\digamma$*|, f|*$\digamma$*|, h|*$\digamma$*|}]
\end{lstlisting}
This solution is effective and simple enough that we did not encounter any need to build more complex structures to describe the Lorentz frame, during the work that led to the results in \cite{Torsello_2019}.\footnote{We could introduce another manifold with another set of indices representing the Lorentz frame, and then connect the two manifolds with a suitable map. However, as we said, there was no need to do that, which would perhaps result in a more mathematically rigorous, but less intuitive structure of the code.}

\subsection{Raising and lowering indices}

In bimetric relativity there are (at least) two metric sectors, therefore raising and lowering indices must be done accordingly. \code{xAct} allows the definition of one active metric only, which automatically raises and lowers all the indices. The active metric in \code{bimEX}, at the present version, is $\gSp$, i.e., the physical spatial part of the metric $\gMet$. This is clearly a problem in bimetric relativity, since we do not want one metric to raise or lower the indices in the other metric sectors.

This means that we need to take care of contractions explicitly. This is done by defining the tensors with some canonical indices and never write them with the indices in different positions. Suppose the user defines the extrinsic curvature in the $\fMet$-sector $\fK_{ij}$, and wants to raise the index $i$. The user should not write it in upper position directly, as in $\fK^i{}_j$, because this expression in \code{xAct} would be equivalent to $\gSp^{ik}\fK_{kj}$. Rather, the user should explicitly write $\fSp^{ik}\fK_{kj}$. Also, \code{bimEX} includes a function, inspired by the built-in \code{xAct} function \code{Simplification}, which simplifies the abstract expressions without raising and lowering indices. It is called \code{SafeSimplification} and it is defined as,
\begin{lstlisting}[caption={Definition of the command \code{SafeSimplification}, which simplifies expressions involving abstract tensors in \code{xAct}, without raising and lowering indices.}]
	SafeSimplification[expr_]:=
	ToCanonical[expr,
		UseMetricOnVBundle |*$\rightarrow$*| None
	] //Simplify
\end{lstlisting}
The explicit writing of every contraction and the use of \code{SafeSimplification} results in unambiguous computations.

\subsection{The definitions of the covariant BSSN equations}

The next step is the definition of the covariant BSSN equations as abstract tensors in \code{xAct}. We take the Hamiltonian constraint in the $\gMet$-sector as an example, but the same procedure is implemented for all the evolution and constraint equations in \cite{Torsello_2019}. At the present stage, the package only includes the covariant BSSN equations, not the standard \nPlusOne ones. However, \code{bimEX} easily allows to implement them, since it is enough to mimic the procedure described below. Hence, in order to be compatible with later versions that could include also other sets of equations, we prefix the objects referring to the covariant BSSN with the string \code{cBSSN\$}. We can see this explicitly in the definition below
\begin{lstlisting}[caption={Definition of the Hamiltonian constraint in the $\gMet$-sector as an abstract tensor used as a placeholder.},label={lst:gHC}]
	DefTensor[cBSSN$gHamiltonianConstraint[],
		{|*$\Sigma$*|t, t, |*$g\digamma,f\digamma,h\digamma$*|}
	]
\end{lstlisting}
This tensor is now only a placeholder for the Hamiltonian constraint in the $\gMet$-sector. We would like the component of this scalar in a given chart to be the Hamiltonian constraint in that chart. On the other hand, we also want to manipulate the abstract Hamiltonian constraint. Therefore, we define an \code{IndexRule} to replace \code{cBSSN\$gHamiltonianConstraint[]} with its explicit tensorial expression in the covariant BSSN. The rule is called \code{Instantiate\$gHC}.
\begin{lstlisting}[caption={Example usage of the \code{IndexRule} to instantiate the Hamiltonian constraint in the $\gMet$-sector to the abstract equation.},label={lst:indexrule}]
	cBSSN$gHamiltonianConstraint[]
	/.Instantiate$gHC
\end{lstlisting}
The command in Listing~\ref{lst:indexrule} prints the abstract Hamiltonian constraint in the covariant BSSN formulation. There are also other \code{IndexRules}: \code{InstantiateConstraints}, \code{InstantiateEvolution} and \code{InstantiatePDE} do the same job for the constraint equations only, for the evolution equations only and for both of them.

Once the equations are instantiated, one can use the \code{xAct} functions to manipulate them. All the bimetric interactions and sources are also defined as abstract tensors by \code{bimEX}.

\section{The concrete equations}
\label{sec:concrete}

\subsection{The computation of the square root in \eqref{eq:R}}
\label{subsec:sqrt}

The computation of $(\sEta^{-1}\sRbar^\tr\sEta\sRbar)^{1/2}$ in \eqref{eq:R} needs to be performed with care, since we are dealing with symbolic manipulation. Hence, the computation can be inefficient or result in a very complicated expression for $(\sEta^{-1}\sRbar^\tr\sEta\sRbar)^{1/2}$, depending on the chosen ansatz and on the algorithm used to compute it.

We have implemented three algorithms to compute this square root. Depending on the ansatz, one can perform better than the others. Which method is the best has to be inspected case by case. The user can choose in a quite simple way which method to use, as we will explain in the next subsection. The first method, which is the default and simplest one, is based on the built-in Mathematica function \code{MatrixPower}. The implementation is the following,
\begin{lstlisting}[emph={Matrix}, caption={Algorithm to compute the square root of a real matrix, using the Mathematica built-in function \code{MatrixPower}.},label={lst:matrixpower}]
	RealMatrixSqrt[Matrix_?MatrixQ] := 
	Assuming[
		Flatten[
			Flatten[
				Simplify[Matrix]
			]
		] |*$\in$*| Reals, 
		Simplify[MatrixPower[Matrix, 1/2]]
	];
\end{lstlisting}
The second method is an adapted version of the one reported in \cite{MathWeb}, and consists in computing the polar decomposition of $\sRbar^{-1}$, using the built-in Mathematica function $\mathtt{SingularValueDecomposition}$,
\begin{lstlisting}[emph={f,f_,U,V,W}, emphstyle=\textit, caption={This function computes the polar decomposition of a generic matrix, by using the built-in Mathematica function \code{SingularValueDecomposition}. Adapted from the algorithm in \cite{MathWeb}.},label={lst:polar}]
	PolarDecomposition[Matrix_?MatrixQ] :=
	Module[{U, W, V},
		{U, W, V} =
		Assuming[
			Flatten[
				Flatten[
					Simplify[Matrix]
				]
			] |*$\in$*| Reals, 
			SingularValueDecomposition[Matrix]
		] //. {Conjugate[f_] |*$:\rightarrow$*| f, Re[f_] |*$:\rightarrow$*| f, Abs[f_] |*$:\rightarrow$*| f};
		Return[
			{U.W.Transpose[U], U.Transpose[V]}
		];
	]
\end{lstlisting}
Note that, in these first two methods, we explicitly use the fact that we are dealing with real quantities.
The third method is the implementation of the algorithm presented in \cite{FRANCA1989459}. This algorithm is very efficient when dealing with numbers. In the case of symbolic manipulation, it is not guaranteed that it can perform better than the other two. Its implementation is,
\begin{lstlisting}[emph={M,S,A,A1,A2,A3,S1,S2,S3,k,l}, emphstyle=\textit, caption={Implementation of the algorithm in \cite{FRANCA1989459} to compute the square root of a positive definite $3\times 3$ matrix.},label={lst:posdef}]
SquareRootOf3DPositiveDefiniteMatrix[
	Araw_?MatrixQ, 
	OptionsPattern[{ApplyFunction |*$\rightarrow$*| Identity}]
] := 
Module[
	{M, S, A, A1, A2, A3, S1, S2, S3, k, l, |*$\phi$*|, |*$\lambda$*|},
	A = Araw // OptionValue[ApplyFunction];
	A1 = Tr[A];
	A2 = (Tr[A]^2 - Tr[A.A])/2;
	A3 = Det[A];
	k = A1^2 - 3 A2;
	If[k == 0,
	
		Print["- k == 0. The square root is a multiple of the identity."];
		S = Sqrt[A1/3] IdentityMatrix[3] // OptionValue[ApplyFunction];,
		
		l = A1 (A1^2 - 9/2 A2) + 27/2 A3;
		|*$\phi$*| = ArcCos[l/k^(3/2)];
		|*$\lambda$*|= (1/3 (A1 + 2 Sqrt[k] Cos[|*$\phi$*|/3]))^(1/2);
		S3 = Sqrt[A3];
		S1 = |*$\lambda$*| + (-|*$\lambda$*|^2 + A1 + (2*S3)/|*$\lambda$*|)^(1/2);
		S2 = (S1^2 - A1)/2;
		S = 1/(S1*S2 - S3) (S1*S3 IdentityMatrix[3] + (S1^2 - S2) A - A.A) // OptionValue[ApplyFunction];,
			
		l = A1 (A1^2 - 9/2 A2) + 27/2 A3;
		|*$\phi$*| = ArcCos[l/k^(3/2)];
		|*$\lambda$*|= (1/3 (A1 + 2 Sqrt[k] Cos[|*$\phi$*|/3]))^(1/2);
		S3 = Sqrt[A3];
		S1 = |*$\lambda$*| + (-|*$\lambda$*|^2 + A1 + (2*S3)/|*$\lambda$*|)^(1/2);
		S2 = (S1^2 - A1)/2;
		S = 1/(S1*S2 - S3) (S1*S3 IdentityMatrix[3] + (S1^2 - S2) A - A.A) // OptionValue[ApplyFunction];
	];
	Return[S];
]
\end{lstlisting}
This function has the option \code{ApplyFunction}, which is discussed in the next subsection.

To reproduce the results in \cite{Kocic:2018ddp} and obtain the results in \cite{Torsello_2019,Torsello_2019b}, we used the algorithm in Listing~\ref{lst:matrixpower}, since the chosen ansatz imply that $\sEta^{-1}\sRbar^\tr\sEta\sRbar$ is diagonal, and the function \code{MatrixPower} is efficient enough. In this case, the algorithm in Listing~\ref{lst:posdef} gives a much more complicated expression, which has to be simplified in a second step to give the simpler expression obtained with the algorithm in Listing~\ref{lst:matrixpower}. Also, in this case the efficiency of the algorithm in Listing~\ref{lst:polar} is comparable with the one in Listing~\ref{lst:matrixpower}.

We stress the fact that, since we are dealing with symbolic manipulation, it is important to recognize repetitive patterns appearing in the equations. For this reason we defined the shifted elementary symmetric polynomials in \cite{doi:10.1063/1.5100027}, which considerably simplify the computations and increase the efficiency of the code. We verified this explicitly in the case of spherical symmetry. Our approach was to blindly compute a part of the decomposition as the first step, and then to recognize repetitive patterns which can be represented in Mathematica as single symbols. The substitution of the repetitive patterns with single symbols speeds up the symbolic manipulation tremendously.

\subsection{Computation of the components in a given chart}
\label{sec:computeDecomposition}

Here we describe the functions that compute the bimetric BSSN decomposition in a given chart. First, one has to define a chart. This is made easier by the function \code{DefChartScalars}, which is based on the built-in \code{xAct} function \code{DefChart} and has its same \code{Options}. It defines a chart on the spacelike hypersurface, assigns components to the identity operator in that chart, and sets the independence of the parameters \code{t,g\digamma,f\digamma,h\digamma} from the coordinates in the chart, i.e.,
\begin{lstlisting}[caption={Setting the independence of the parameter from the spatial coordinates.}]
	PDOfBasis[ChartName][i_][t] := 0;
	PDOfBasis[ChartName][i_][g|*$\digamma$*|] := 0;
	PDOfBasis[ChartName][i_][f|*$\digamma$*|] := 0;
	PDOfBasis[ChartName][i_][h|*$\digamma$*|] := 0;
\end{lstlisting}
In addition, \code{DefChartScalars} sets the values of the variables \code{FirstChart} and \code{DefaultChart} to the name of the chart, henceforth considered as the default chart for all the functions. The variable \code{DefaultChart} can be modified, and in the case of multiple charts it is reset to the last defined chart. The value of \code{FirstChart} is \code{Protected} instead. In this way, \code{bimEX} can handle multiple charts simultaneously. In case the user is using only one chart, there is no need to specify it when dealing with the components of the tensors. \code{DefChartScalars} has the option \code{ChartAssumptions}, which allows to specify a list containing the assumptions on the coordinates. As an explicit example, this commands defines the spherical polar chart on the spacelike hypersurface,
\begin{lstlisting}[caption={Example definition of the spherical polar chart.}]
	DefChartScalars[|*$S$*|, {r[], |*$\Theta$*|[], |*$\Phi$*|[]}, 
		ChartAssumptions |*$\rightarrow$*| {r[]>0, |*$\Theta$*|[]>0, |*$\Phi$*|[]>0},
		ChartColor |*$\rightarrow$*| Purple
	]
\end{lstlisting}

Once a chart is defined, the ansatz should be chosen. The ansatz is the input to the function \code{ComputeBSSNDecomposition}. We must choose an ansatz for the primary variables in \eqref{eq:primaryvariables}, plus a background metric for $\hSpBS_{ij}$:
\begin{enumerate}
	\item The conformal factors $\gconf$, $\fconf$
	\item The conformal vielbeins $\gEBS^\textbf{a}{}_i$, $\fEtrBS^\textbf{a}{}_i$
	\item The vector $\sLp^\textbf{a}$ which completely determines $\sLs^\textbf{a}{}_\textbf{b}$ and, together with $\gEBS^\textbf{a}{}_i$, $\fEtrBS^\textbf{a}{}_i$, $\sRs^\textbf{a}{}_\textbf{b}$ (hence $L$)
	\item The shift vector $\hShift^i$ of the geometric mean metric $\hMet$
	\item The conformal extrinsic curvatures $\gABS^i{}_j$, $\fABS^i{}_j$
	\item The conformal connections $\gL^i$, $\fL^i$
	\item The three background metrics for $\gSpBS_{ij}$, $\fSpBS_{ij}$, $\hSpBS_{ij}$
\end{enumerate}
\code{ComputeBSSNDecomposition} uses background connections arising from background metrics. The background metric for $\hSpBS_{ij}$ is an input variable needed to compute the dynamics of the geometric mean in the covariant BSSN formulation, included for forward compatibility.

After we choose the ansatz for these variables, we give it to \code{ComputeBSSNDecomposition} as input. This function computes all the components of the bimetric interactions and sources, storing them both as the components of the abstract tensors in \code{xAct}, and as plain lists in \code{Mathematica}. Hence, the user will be able to use the computed components within \code{bimEX} or not, as we will discuss in Sec.~\ref{subsec:export}. Since the components are saved as plain lists, the user can extend \code{bimEX} to handle them in \code{CTensor} as well, in a straightforward way. All geometrical quantities---Ricci tensor, Christoffel symbols, etc.---in the metric sectors are also computed by \code{ComputeBSSNDecomposition}, making use of the \code{xAct} built-in function \code{MetricCompute}. \code{ComputeBSSNDecomposition} also performs many internal checks to verify some of the relations involving the bimetric interactions. If any of these checks fails, the evaluation aborts and an informative message about the error is printed.

The function \code{ComputeBSSNDecomposition} has five options that can make the computations more efficient, and allow the user to decide what to compute. The options are,
\begin{enumerate}
	\item \code{ChartName}. The default value for this option is \code{DefaultChart}. If the user needs to use multiple charts, they should set \code{ChartName} to the name of the chart they want to compute the decomposition in.
	\item \code{ComputeGeometryOf}. This option should be set equal to a list containing the names of the metrics whose geometrical quantities we want to compute. The default value is $\left\lbrace \gSp,\fSp \right\rbrace$, which means that the geometries of $\gSp,\gSpBS$ and $\fSp,\fSpBS$ will be computed, but the geometry of $\hSp,\hSpBS$ will not, since it slows down the execution. The user can decide to compute the geometries of any of the metrics.
	\item \code{ApplyFunction}. This option allows the user to specify a function to be applied to the results of the computations performed in \code{ComputeBSSNDecomposition}. The default value is the identity, i.e., nothing is applied to the computed expressions. The chosen function has a direct impact on the efficiency of \code{ComputeBSSNDecomposition}. It is desirable to identify recurrent patterns in the equations to make the computation faster and more useful. The user can define a function $f$ that recognizes these patterns in the chosen ansatz, and set $f$ as the value of \code{ApplyFunction}. This increased the efficiency tremendously for our computations in spherical symmetry.
	\item \code{SqrtAlgorithm}. This option allows the user to choose the algorithm to compute the square root matrix $(\sEta^{-1}\sRbar^\tr\sEta\sRbar)^{1/2}$. It can be set to three strings, \code{``MatSqrt"}, \code{``PolDec"}, \code{``PosDefSqrt"}. They refer to the three algorithm discussed in Sec.~\ref{subsec:sqrt}, respectively to the algorithms in Listing~\ref{lst:matrixpower}, Listing~\ref{lst:polar} and Listing~\ref{lst:posdef}.
	\item \code{IndVariables}. This option allows the user to specify a list containing the independent variables to be set as arguments for the primary fields. The default value is the string \code{``AutoDetect"}, which makes \code{ComputeBSSNDecomposition} to inspect the ansatz and find the independent variables. The following shortcut is then defined, for a generic function $f$,
$$\underline{f\textunderscore }\coloneqq f \mathtt{[\left\langle detected\; or\; specified\; variables\right\rangle]}.$$
\end{enumerate}

An example call of \code{ComputeBSSNDecomposition} is
\begin{lstlisting}[caption={Example call of \code{ComputeBSSNDecomposition}. Here, the symbol $\langle X\rangle$ represents the components of the object $X$ in the chosen ansatz.}, label={lst:computedecomposition}]
	ComputeBSSNDecomposition[
		|*$\left\langle\gconf\right\rangle$*|, |*$\left\langle\fconf\right\rangle$*|,
		|*$\left\langle\gEBS^\textbf{a}{}_i\right\rangle$*|, |*$\left\langle\fEtrBS^\textbf{a}{}_i\right\rangle$*|, |*$\left\langle\sLp^\textbf{a}\right\rangle$*|, |*$\left\langle\hShift^i\right\rangle$*|,
		|*$\left\langle\gABS^i{}_j\right\rangle$*|, |*$\left\langle\fABS^i{}_j\right\rangle$*|, |*$\left\langle\gL^i\right\rangle$*|, |*$\left\langle\fL^i\right\rangle$*|, 
		|*$\left\langle{\gSp_{\mathtt{B}}}_{ij}\right\rangle$*|,  |*$\left\langle{\fSp_{\mathtt{B}}}_{ij}\right\rangle$*|,  |*$\left\langle{\hSp_{\mathtt{B}}}_{ij}\right\rangle$*|,
		ApplyFunction |*$\rightarrow\left\langle\mbox{desired function}\right\rangle$*|,
		ComputeGeometryOf |*$\rightarrow\cb{\langle\mbox{desired metrics}\rangle}$*|
  ]
\end{lstlisting}
Note that the order of the input variables in \code{ComputeBSSNDecomposition} is important, and that the two conformal vielbeins $\gEBS,\fEtrBS$ have to be upper triangular.

We remark the following limitation. The \code{xAct} function \code{MetricCompute}, used by \code{ComputeBSSNDecomposition} to compute the components of the geometrical objects associated with the metrics, does not store the computed components in all the charts. For example, suppose the user uses two charts $C_1$ and $C_2$. The user executes \code{MetricCompute} in the chart $C_1$ first, and immediately afterwards in the chart $C_2$. The second execution in the chart $C_2$ \emph{overwrites} the previously computed components in the chart $C_1$. This has to be taken into account when using multiple charts in \code{bimEX}.

\subsection{How to access to the computed decomposition besides \code{xAct}}
\label{subsec:lists}

There is a standard notation for the lists storing the components of the tensors computed by \code{ComputeBSSNDecomposition}, which helps the user to remember their names. For definiteness, let's consider the components of the metric $\fSpBS_{ij}$ and the conformal extrinsic curvature $\gABS_{ij}$. They are stored in the lists defined as,
\begin{lstlisting}[caption={Definitions of the lists containing the components of $\gSpBS_{ij}$ and $\gABS_{ij}$, respectively.},label={lst:lists}]
	|*$\varphi$*|c|*$ \blacksquare$*|[]:= |*$\varphi$*|c|*$ \blacksquare$*|[DefaultChart]
	gA|*$\blacktriangledown\blacktriangledown$*|[]:= gA|*$\blacktriangledown\blacktriangledown$*|[DefaultChart]
\end{lstlisting}
where the \qm{c} in \code{\varphi c} stands for \qm{conformal} metric. If the user uses one chart only, there is no need to specify the name of the chart. The symbols in the names tell us what type of components are we looking at. The symbol $\blacktriangledown\blacktriangledown$ means that the list \code{gA}$\blacktriangledown\blacktriangledown$\code{[]} stores the components of $\gABS_{ij}$ with both lower indices. The components of $\gABS^i{}_j$ are stored in the list \code{gA}$\blacktriangleup\blacktriangledown$\code{[]}, and so on. This holds for all the rank 2 tensors defined in \code{bimEX}. However, there are exceptions to this rule. One exception, as we see in Listing~\ref{lst:lists}, concerns the metrics. The list of components of $\fSpBS$ contains the symbol $\blacksquare$ in its name. The component of the inverse conformal metric are stored in the list named $\overset{-1}{\varphi c} \blacksquare$\code{[]}. The mixed components of the metrics are those of the identity, and are stored in lists named with the $\blacktriangleup\blacktriangledown$ and $\blacktriangledown\blacktriangleup$ symbols. Another exception is given by the lists containing the components of Lorentz linear operator, e.g., $\sRs,\sLs$, and the vielbeins. These lists have a rectangle in their name, given by the Mathematica code \code{\backslash[FilledRectangle]}.

Now the user can access the components stored in plain lists. Suppose the user computes the decomposition and wants to see it. The function \code{PrintDecomposition} prints the components of the main variables in the decomposition. The user can call it as follows,
\begin{lstlisting}[caption={An example call to the function \code{PrintDecomposition}.}, label={lst:printd}]
	PrintDecomposition[
		Of |*$\rightarrow$*| {g, f, h}, 
		ApplyFunction |*$\rightarrow\left\langle\mbox{desired function}\right\rangle$*|
	]
\end{lstlisting}
\code{PrintDecomposition} has no arguments and two options, \code{Of} and \code{ApplyFunction}. The value of \code{Of} is a list containing the names of the sectors which we want to print the decomposition of. In Listing~\ref{lst:printd}, the command will print the quantities associated with all three metric sectors.

\subsection{From abstract to concrete equations in \code{xAct}}
\label{sec:toConcrete}

Once \code{ComputeBSSNDecomposition} is executed, all the bimetric interactions, sources and the geometrical quantities of the metrics have assigned components in \code{xAct}. The function \code{ToConcrete} takes an abstract equation, and writes it in components in the given chart. Its efficiency depends on the level of instantiation that the user can control via three boolean options, listed below and bespoke for the covariant BSSN equations,
\begin{enumerate}
	\item \code{ConcreteSources}. If \code{True}, it will instantiate the bimetric sources. Setting this options to \code{False} can be very useful for the readability of the final equations, and their manipulation in Mathematica, since there are less terms to manipulate. These terms do not contain derivatives of the dynamical fields.
	\item \code{ConcreteRicci}. If \code{True}, it will instantiate the Ricci tensors in
	\begin{subequations}
	\label{eq:Ricciidentity}
		\begin{align}
			\gRBS_{ij}								&\coloneqq -\dfrac{1}{2}\gSpBS^{k\ell}\gDback_{k}\gDback_{\ell}\gSpBS_{ij}+\gSpBS_{k(i}\gDback_{j)}\gL^k \nonumber \\
														&\quad\ -\gSpBS^{k\ell}\gSpBS_{m(i}\gRback_{j)k\ell}{}^m+\gSpBS^{\ell m}\gDG^k_{\ell m}\gDG_{(ij)k} \nonumber \\
														&\quad\ +\gSpBS^{k\ell}\rb{2\gDG^m_{k(i}\gDG_{j)m\ell}+\gDG^m_{ik}\gDG_{mj\ell}}, \\
			\fRBS_{ij}								&\coloneqq -\dfrac{1}{2}\fSpBS^{k\ell}\fDback_{k}\fDback_{\ell}\fSpBS_{ij}+\fSpBS_{k(i}\fDback_{j)}\fL^k \nonumber \\
														&\quad\ -\fSpBS^{k\ell}\fSpBS_{m(i}\fRback_{j)k\ell}{}^m+\fSpBS^{\ell m}\fDG^k_{\ell m}\fDG_{(ij)k} \nonumber \\
														&\quad\ +\fSpBS^{k\ell}\rb{2\fDG^m_{k(i}\fDG_{j)m\ell}+\fDG^m_{ik}\fDG_{mj\ell}},
		\end{align}	
	\end{subequations}
	appearing in the evolution equations for the conformal extrinsic curvatures $\gABS^i{}_{j},\fABS^i{}_{j}$ \cite{PhysRevD.79.104029,Torsello_2019}. Turning this option to \code{False} speeds up the computation considerably, and helps to see the structure of the equations, which is needed to optimize them before proceeding to the numerical integration.
	\item \code{ConcreteShift}. If \code{True}, it will instantiate the shift vector of $\gMet$ and $\fMet$ according to \eqref{eq:shiftsymcond}. This has the same advantages of \code{ConcreteSources}.
\end{enumerate}
The default value for the three options is \code{False}. \code{ChartName} is also an option for \code{ToConcrete}, and works in the same way. Setting the three options to \code{False} results in a very efficient instantiation algorithm.

Three more functions are defined to be able to instantiate the bimetric sources, the Ricci tensors and the shifts independently from \code{ToConcrete}. They are called \code{ToConcreteSources}, \code{ToConcreteRicci} and \code{ToConcreteShift}. They can be applied to a concrete equation to instantiate the desired components.

Note that, even if we do not instantiate the components of any of these quantities by setting the three options of \code{ComputeBSSNDecomposition} to \code{False}, only their non-zero components will be kept uninstantiated. The zero components will be set to zero not to keep unnecessary terms and simplify the output. The recognition of the zero components is done by \code{ComputeBSSNDecomposition}. For definiteness, let's consider the Ricci tensor as an example. \code{ComputeBSSNDecomposition} computes explicitly all the components of the Ricci tensor, and recognizes the zero ones. Then, it will set to zero all the respective components of the abstract Ricci tensor in \code{xAct}. The non-zero components of the abstract Ricci tensor in \code{xAct} will be set equal to some (appropriately named) scalar functions, acting as placeholders. The functions \code{ToConcrete}, \code{ToConcreteSources}, \code{ToConcreteRicci} and \code{ToConcreteShift} only \emph{replace} the placeholders scalar functions with the actual components computed only once by \code{ComputeBSSNDecomposition}. This results in a fast execution for all the \code{ToConcrete} commands.

The function \code{ToConcrete} is a switch function which depends on the values of the three options of instantiation. It calls different functions depending on the values of its options. Suppose we choose to set the three options to False. Then, \code{ToConcrete} calls only the function \code{ToConcrete\$Basic}. The function \code{ToConcrete\$Basic} is based on the built-in \code{xAct} functions---\code{ToBasis}, \code{TraceBasisDummy}, \code{ComponentArray}, \code{ToValues}, \code{InChart}---but it implements other algorithms as well, since those functions alone do not suffice to write everything in components.\footnote{As an example, the abstract coordinate derivative \code{PD}, used by \code{xAct} in the abstract equations, has to be replaced by hand by the partial derivative in the given basis \code{PDOfBasis}. The Christoffel symbols of the metrics need to be taken into account separately as well.} Essentially, \code{ToConcrete\$Basic} instantiates everything but the Ricci tensors, the bimetric sources and the shift vectors. Suppose that we want to instantiate an abstract equation including the shifts. We set the option \code{ConcreteShift} in \code{ToConcrete} to \code{True}. \code{ToConcrete} then calls \code{ToConcrete\$Basic} first, and \code{ToConcreteShift} later. The same applies for all the other options' combinations.

As an example, the code to instantiate the Hamiltonian constraint in the $\gMet$-sector in the \code{DefaultChart} is
\begin{lstlisting}[caption={Example of the use of the \code{ToConcrete}. We first instantiate the Hamiltonian constraint in the $\gMet$-sector to the abstract equation with \code{Instantiate\$gHC}, and then apply \code{ToConcrete} to it. The output is the Hamiltonian constraint in components in the chart given by \code{DefaultChart}.},label={lst:usetoconcrete}]
	cBSSN$gHamiltonianConstraint[]
	/.Instantiate$gHC //ToConcrete
\end{lstlisting}

\subsection{Exportation of the bimetric decomposition and equations}
\label{subsec:export}

At this point, we know how to compute the BSSN decomposition and how to write the abstract equations in components, with the desired level of instantiation. In the case of spherical symmetry, on a HP Z240 SFF Workstation, with an Intel(R) Core(TM) i7-6700 CPU \@ 3.40GHz and 16.0 GB of RAM memory, running Windows 8 64-bit, \code{Mathematica} 11.0 and \code{xAct} 1.1.3, the execution of \code{ComputeBSSNDecomposition}---computing the geometrical quantities for all the six metrics---takes about 2 minutes, and the instantiation and simplification of all the bimetric covariant BSSN equations takes about 10 minutes in total. Once the decomposition is computed and the equations are instantiated and simplified to the desired level, they can be exported into an .m file. This removes the need to make the same computations each time that the user needs to use, e.g., the constraint equations in spherical symmetry.

We already said that \code{ComputeBSSNDecomposition} saves the computed components of the tensors both as lists and as components stored in the \code{xAct} tensors. This is because we do not want to be necessarily constrained to the \code{xAct} bundle. The user should be able to export the decomposition in a file which does not have any memory of \code{xAct}. In this case, the user could open a Mathematica notebook, load the .m file where the decomposition is saved in the form of plain lists, and use these lists to perform the desired computations using only the standard Mathematica functions. On the other hand, the user also needs to be able to export the decomposition within \code{xAct}, i.e., to export all the definition and rules defined by \code{bimEX} in \code{xAct}. In this second case, the user should be able open a Mathematica notebook, load \code{bimEX} and the .m file where the decomposition in \code{xAct} is exported, and use \code{bimEX} and the \code{xAct} bundle to perform the desired computations.

Before introducing the two functions which export the bimetric decomposition and equations in the two different ways outlined above, we need to hightlight another feature of \code{ComputeBSSNDecomposition}. This function does \emph{not} assign components to the abstract tensors representing the covariant BSSN constraint and evolution equations. For clarity, it does not assign components to \code{cBSSN\$gHamiltonianConstraint[]} in Listing~\ref{lst:gHC}. This is the case for two reasons. First, \code{ComputeBSSNDecomposition} performs all computations \code{ToConcrete} needs in order to write the equations in components. Therefore, for \code{ComputeBSSNDecomposition} to assign components to the abstract equations, \code{ToConcrete} would be needed to be part of it. On the other hand, \code{ToConcrete} needs \code{ComputeBSSNDecomposition} to be executed first, so this type of implementation would require a basic restructuring of the code. Second, after the execution of \code{ToConcrete}, the equations are in concrete form, but they are not necessarily written in the simplest possible form. Hence, one may want to simplify or manipulate them, and then set the simplified equations as the components of the abstract tensors (which are the ones being exported).

The components are assigned to the tensors representing the covariant BSSN equations by the function \code{AssignComponents}, which has \code{ChartName} and \code{ApplyFunction} as options. It takes as arguments two ordered lists, the first one containing the abstract tensors to which we want to assign the components, and the second one containing the components. It applies the value of \code{ApplyFunction} to the list of components, and set them as the components of the abstract tensors in the chart \code{ChartName}. Again, the default value of \code{ChartName} is \code{DefaultChart}, and the default value of \code{ApplyFunction} is the identity. At the present stage, this function works for tensors up to rank 2. After having assigned the components, the user can access the concrete equations with the following code, as an alternative to the one in Listing~\ref{lst:usetoconcrete},
\begin{lstlisting}[caption={After having assigned the components to the abstract tensors representing the abstract equations, we can directly apply \code{ToConcrete} to them. The output is the Hamiltonian constraint in components in the chart given by \code{DefaultChart}.}]
	cBSSN$gHamiltonianConstraint[] //ToConcrete
\end{lstlisting}

We are now ready to export the decomposition to an .m file. There are two functions to do that, \code{ExportDecomposition} and \code{ExportDecomposition\$xAct}. The first one exports all the components stored into lists, in an .m file which is independent from the \code{xAct} bundle and from \code{bimEX} as well. Its arguments are the fields in the chosen ansatz, without the conformal factors. It has four options, \code{ChartName}, \code{ApplyFunction}, \code{NameFile} and \code{OtherQuantities}. The first two work in the same way as for the other functions, and \code{NameFile} allows to specify the name of the exported .m file. The default value of \code{NameFile} is \code{``Bimetric\_decomposition\_\langle time\; and\; date\rangle .m"}. The option \code{OtherQuantities} allows the user to add other quantities to be exported. It should be set to a list of two lists. The first list includes the names of the variables to be exported, and the second list the expressions to be saved into these variables. This makes it possible to export the concrete equations, which are not exported by \code{ExportDecomposition} by default. If the option \code{OtherQuantitites} is used, it is necessary to execute the command \code{MapThread[Set, {\$\$names, \$\$values}]} after loading the .m file. In addition, the expressions included in the value of \code{OtherQuantities} and specified by the user, must be totally instantiated, i.e., they should not contain any reference to xAct. \code{ExportDecomposition} locally clears the \code{UpValues} of the variables and uses \code{DumpSave} to export them, since this function does not export the \code{FullDefinition} of a variable, contrary to \code{Save}. The variables exported by default are named with the same convention described in \autoref{subsec:lists}, but the last two characters in the names, namely \code{[]}, are replaced by \code{\$}. For clarity, \code{gA}$\blacktriangleup\blacktriangledown$\code{[]} becomes \code{gA}$\blacktriangleup\blacktriangledown\$ $ in the exported .m file.

\code{ExportDecomposition\$xAct} takes no arguments and exports all the information about the tensors defined in \code{xAct} and their components into an .m file, using \code{Save}. Before loading the .m file, the user needs to load \code{bimEX}, since it defines the spacelike hypersurface and all the abstract tensors. The .m file exported by \code{ExportDecomposition\$xAct} will then \emph{overwrite} all the information stored in it to the basic information defined by \code{bimEX}. One could also think about this as \emph{adding} the information concerning the components of all the tensors within \code{xAct}, to the basic information contained in \code{bimEX}. \code{ExportDecomposition\$xAct} only has the option \code{NameFile}, whose default value is \code{``Bimetric\_decomposition\_xAct\_\langle time\; and\; date\rangle .m"}.

\section{Conclusions}
\label{sec:conclusions}

We presented \code{bimEX}, a package to perform bimetric exact computations in the \threePlusOne formalism in \code{Mathematica}. It makes extensive use of the \code{xAct} bundle and can handle both abstract and concrete---i.e., with components---computations.

The package was written during the work to obtain the results in \cite{Torsello_2019,Torsello_2019b}. Therefore, at the present stage, it includes the covariant BSSN equations written as abstract tensors, together with the bimetric \threePlusOne and BSSN decompositions \cite{Kocic:2018ddp,Torsello_2019}. The user can manipulate these equations abstractly by using the \code{xAct} built-in functions. However, there is a key factor that must be reminded: the user must have complete control on how the indices are raised and lowered, since the objects in one metric sector cannot be contracted with the other metric.
This constitute a problem in \code{xAct}, because it does not allow to define only frozen metrics, where by frozen metrics we mean metrics that do not raise and lower indices automatically. Our solution to this problem is to use the \code{xAct} function \code{ToCanonical} always with the option \code{UseMetricOnVBundle\rightarrow None}; we defined the function \code{SafeSimplification} which automatically does this. The function \code{SafeSimplification} never raises and lowers indices during the simplifications. In addition, one has to write contractions explicitly.

The package allows to choose an ansatz on the primary variables of the theory, listed in \eqref{eq:primaryvariables}, to give it as input to the function \code{ComputeBSSNDecomposition}, and to obtain the bimetric decomposition as output. After the bimetric decomposition is computed, the function \code{ToConcrete}, applied to an abstract quantity, writes it into components. This function allows the user to choose the level of instantiation of the output, in order to increase efficiency and readability of the equations written in components.

The package allows the user to print the decomposition via the function \code{PrintDecomposition}, and export it, via the functions \code{ExportDecomposition} and \code{ExportDecomposition\$xAct}. The first function exports the decomposition into a .m file which is independent from the \code{xAct} bundle and \code{bimEX}, since all the components of the tensors are saved into lists---i.e., matrices. The second function exports all the definitions of the \code{xAct} tensors into an .m file, which requires \code{bimEX} (hence \code{xAct}) to be loaded first, in order to work properly. The user is then able to choose how to work with the decomposition. 

In the computation of the bimetric decomposition, the square root of the positive definite $3\times 3$ matrix $\sEta^{-1}\sRbar^\tr\sEta\sRbar$, which originates from the solution to the symmetrization condition \eqref{eq:spsymcond}, has to be computed. Three algorithms are implemented for this, one taken from \cite{FRANCA1989459}, and the other two relying on Mathematica built-in functions. The user decides which algorithm to use, via an option of \code{ComputeBSSNDecomposition}, which computes the square root. Since we deal with symbolic manipulation, there is no general rule about which method to use. The efficiency depends on the chosen ansatz and on the user-defined simplification functions. We suggest to define simplification functions which recognize repetitive patterns into the expressions and write them as single symbols.

We stress that this package can compute the covariant BSSN equations for \emph{any} desired ansatz. It has been tested extensively for the spherically symmetric case, and should be useful to anyone working on numerical bimetric relativity.

\section*{Acknowledgments}

We are thankful to Mikica Kocic for all the shared and fruitful Mathematica sessions, and to Edvard M\"{o}rtsell and Mikica Kocic for reading the paper and providing useful comments.

\appendix



\bibliographystyle{elsarticle-num}
\bibliography{biblio}







\end{document}
